\documentclass[conference]{IEEEtran}
\IEEEoverridecommandlockouts
\usepackage{cite}
\usepackage{amsmath,amssymb,amsfonts}
\usepackage{algpseudocode,algorithm,algorithmicx}
\usepackage[capitalise]{cleveref}
\usepackage{graphicx}
\usepackage{textcomp}
\usepackage{xcolor}
\usepackage{makecell}
\def\BibTeX{{\rm B\kern-.05em{\sc i\kern-.025em b}\kern-.08em
T\kern-.1667em\lower.7ex\hbox{E}\kern-.125emX}}
\newcommand{\etal}{\emph{et al. }}

\newtheorem{theorem}{Theorem}

\newtheorem{proposition}[theorem]{Proposition}

\newtheorem{remark}{Remark}

\begin{document}

\title{Ultra-Reliable and Low-Latency Short-Packet Communications for Multihop MIMO Relaying\\
\thanks{This research was supported, in part, by Basic Science Research Program through the National Research Foundation of Korea (NRF) funded by the Ministry of Education (NRF-2019R1A6A1A03032119) and by the NRF grant funded by the Korea government (MSIT) (NRF-2019R1F1A1061934).}
}
\author{\IEEEauthorblockN{Ngo~Hoang~Tu}
\IEEEauthorblockA{\textit{Department of Smart Energy Systems} \\
	\textit{Seoul National University of Science and Technology}\\
	Seoul, Republic of Korea \\
	ngohoangtu@seoultech.ac.kr}
\and
\IEEEauthorblockN{Kyungchun~Lee}
\IEEEauthorblockA{\textit{Department of Electrical and Information Engineering and}\\
	\textit{the Research Center for Electrical and Information Technology} \\
	\textit{Seoul National University of Science and Technology}\\
	Seoul, Republic of Korea\\
	kclee@seoultech.ac.kr}
}
\maketitle

\begin{abstract}
This work considers the multihop multiple-input multiple-output relay network under short-packet communications to facilitate not only ultra-reliability but also low-latency communications.
We assume that the transmit antenna selection (TAS) scheme is utilized at the transmit side, whereas either selection combining (SC) or maximum ratio combining (MRC) is leveraged at the receive side to achieve diversity gains.
For quasi-static Rayleigh fading channels and the finite-blocklength regime, we derive the approximate closed-form expressions of the end-to-end (e2e) block error rate (BLER) for both the TAS/MRC and TAS/SC schemes.
The asymptotic performance in the high signal-to-noise ratio regime is derived, from which the comparison of TAS/MRC and TAS/SC schemes in terms of the diversity order, e2e BLER loss, and SNR gap is provided.
The e2e latency and throughputs are also analyzed for the considered schemes.
The correctness of our analysis is confirmed via Monte Carlo simulations.
\end{abstract}
\begin{IEEEkeywords}
Short-packet communication, 
ultra-reliable and low-latency communications,
multihop relaying,
multiple-input multiple-output,
transmit and receive diversity.
\end{IEEEkeywords}

\section{Introduction}
Fifth-generation (5G) networks will reinforce various foremost services, including massive Machine-Type Communications (mMTCs), enhanced Mobile Broadband (eMBB), and ultra-Reliable and Low-Latency Communications (uRLLCs) \cite{navarro2020survey}.
In particular, based on 5G ITU-R and IMT-2020, the 5G networks will fulfill the stringent requirements from (\textit{i}) eMBB with one hundred times the network energy efficiency and three times the spectrum efficiency of IMT-Advanced, a traffic capacity of $10-100$~Mbps/m$^2$, user experienced data rates of 100~Mbps, and peak data rates of $10-20$~Gbps;
(\textit{ii}) mMTCs with the connection density of $10^6$ devices/km$^2$; and
(\textit{iii}) uRLLC with very low latency of around $1-10$~ms \cite{lopez2017ultrareliable}.
From the standpoint of uRLLC, the system needs to satisfy not only low latency, but also ultra-reliability, which refers to packet error rates lower than $10^{-5}$.
Motivated by this, in \cite{polyanskiy2010channel}, the fundamental short-packet communication (SPC) conception has been pioneeringly explored to meet the uRLLC requirements in or even beyond 5G \cite{dao2021survey} by shortening the frame length to as short as 100 channel uses (CUs), in contrast to the conventional systems with long blocklengths.
Specifically, in \cite{polyanskiy2010channel}, Polyanskiy \etal have contributed to obtaining the approximation of the maximum coding rate under the finite-blocklength (FBL) regime.

Besides that, the multiple-input multiple-output (MIMO) system can provide diversity gains to significantly improve reliability, whereas the multihop relay network is known to be an efficient solution to extend coverage, reduce transmit power, and improve reliability for data transmission over short distances.
After the pioneering work in \cite{polyanskiy2010channel}, SPC has recently received much attention, especially in state-of-the-art integration frameworks between SPC and MIMO/relaying system.
In particular, Durisi \etal \cite{durisi2015short} considered a MIMO system for SPC and revealed that the traditional infinite-blocklength metrics would no longer accurately provide estimations for the maximum coding rate in the short-packet-size scenario.
In \cite{ostman2019short} and \cite{zeng2020enabling}, massive MIMO links supporting uRLLC transmissions were successfully leveraged by SPC.
Meanwhile, \cite{yu2020secure} studied the physical-layer secrecy throughput in an uplink massive multi-user MIMO system using short-packet transmission.
Subsequently, in \cite{wei2021secrecy}, a full-duplex multi-user MIMO SPC network was considered in terms of secrecy throughput performance.
Furthermore, \cite{tran2020short} investigated MIMO systems in SPC along with non-orthogonal multiple access support to achieve the benefits of reliability, connectivity, spectral efficiency, and low latency.
From the multihop relay network perspective under SPC, Li \etal \cite{li2020multi} demonstrated that end-to-end (e2e) fountain coding is more suitable for multihop SPCs than random linear network coding. 
Meanwhile, Makki \etal \cite{makki2019end} analyzed the e2e delay and throughput of multihop SPCs.

Although the various benefits of MIMO and multihop relay networks can be provided to SPC for uRLLCs, the comprehensive integrated framework of these schemes under SPC has not been defined.
This paper first investigates the multihop MIMO relay network gains along with SPC to support uRLLCs.
Our main contributions are summarized as follows:
\begin{itemize}
    \item Typical diversity schemes, including transmit antenna selection (TAS), maximum ratio combining (MRC), and selection combining (SC), are utilized to exploit the full diversity gains of our proposed systems.
    \item Over quasi-static Rayleigh fading channels and FBL context, the e2e block error rate (BLER) performance is analyzed, from which the e2e latency and throughput are also investigated.
    \item Based on the asymptotic analysis in the high-SNR region, qualitative conclusions about the diversity orders, e2e BLER loss, and SNR gap between the TAS/MRC and TAS/SC schemes are made.
    \item Our theoretical analysis is verified by Monte Carlo simulations, which show that the e2e performance obtained via theoretical analysis agrees with that via simulation.
    \item \textcolor{black}{The results confirm the benefits from the employment of MIMO and multihop relay network for SPC to meet the uRLLC requirements.}
    
\end{itemize}


\section{System Model}
\label{Sect:SystemModel}


We consider a multihop network under the SPC scheme with $K$ selective decode-and-forward (SDF) relays named ${R_1},{R_2},...,{R_K}$, which assist the communication between one source terminal $\left( {{R_0}} \right)$ and one destination $\left( {{R_{K+1}}} \right)$ over Rayleigh fading channels.
We assume that there is no direct link between the source and destination. 
The MIMO system under consideration consists of $N_T$ transmit antennas and $N_R$ receive antennas, which are installed in each relay node, whereas the source and destination nodes are equipped with only $N_T$ transmit and $N_R$ receive antennas, respectively.
When the $i$th transmit antenna of the node $R_{k-1}$ is chosen for transmission by the TAS scheme, the received signal at the $j$th antenna of the node $R_k$ can be represented as
\begin{align}
\label{Eq:ReceivedSignal}
y_k^{\left( {i,j} \right)} = \sqrt {{P_{k - 1}}} h_k^{\left( {i,j} \right)}x + n_k^{\left( {i,j} \right)},
\end{align}
where $P_{k-1}$ is the transmit power of the node $R_{k-1}$,
$h_k^{\left( {i,j} \right)}$ is the fading coefficient from the $i$th transmit antenna to the $j$th receive antenna at the $k$th hop (${k = \overline {1,K + 1} }$, $i = \overline {1,{N_T}}$, and $j = \overline {1,{N_R}}$)\footnote{The value of $k = \overline {1,K + 1} $, $i = \overline {1,{N_T}}$, and $j = \overline {1,{N_R}}$  will be used throughout this paper if we do not specifically mention.}.
$x$ is a scalar complex baseband transmitted signal with zero mean and unit variance \cite{yang2012cascaded}, i.e., 
${\mathbb{E}}\left\{ {{{\left| x \right|}^2}} \right\} = 1$,
and $n_k^{\left( {i,j} \right)}$ is a zero-mean complex Gaussian noise signal with variance ${\cal N}_0$.

We note that the TAS technique at the transmit side of each hop can achieve a significant reduction in power consumption and hardware cost because only a single radio-frequency (RF) chain is used.
It is assumed that the receiver at each hop has the perfect channel-state information (CSI).
At the transmit side, based on feedback information to the transmitter from the receiver, known as partial CSI feedback \cite{al2019robust}, the index of the best transmit antenna, which maximizes the received SNR, can be determined.
Eventually, the transmitter employs only a single optimal antenna out of $N_T$ antennas to transmit data.
Meanwhile, the MRC and SC are considered at the receive side to achieve receive diversity at each hop.

For TAS/MRC, to achieve the spatial diversity gain at the receive side,
the MRC scheme can be used \cite{liu2020performance}, where the received signals from each channel are coherently combined.
Therefore, the output SNR of the MRC combiner increases with the number of diversity branches.
Subsequently, when we apply TAS and MRC simultaneously for each hop, the instantaneous SNR at the $k$th hop can be written as
$\gamma _k^{{\rm{TAS/MRC}}} = \mathop {\max }\limits_{i = 1,2,...,{N_T}} \sum\limits_{j = 1}^{{N_R}} {\gamma _k^{\left( {i,j} \right)}}$.
Accordingly, the cumulative distribution function (CDF) of $\gamma _k^{{\rm{TAS/MRC}}}$ can be obtained as
\begin{align}
\label{Eq:CDF_TAS/MRC}
{F_{\gamma _k^{{\rm{TAS/MRC}}}}}\left( \gamma  \right) 
= {\left\{ {1 - \exp \left( { - \frac{\gamma }{{{{\bar \gamma }_k}}}} \right)\sum\limits_{n = 0}^{{N_R} - 1} {\frac{1}{{n!}}{{\left( {\frac{\gamma }{{{{\bar \gamma }_k}}}} \right)}^n}} } \right\}^{{N_T}}}.
\end{align}

For TAS/SC, when the SC scheme is employed at the receive side of each hop, the link with the highest received SNR is selected to perform the detection process \cite{yang2012transmit}.
Once this approach is leveraged along with the TAS scheme, a single transmit antenna out of $N_T$ antennas and a single receive antenna out of $N_R$ antennas are jointly chosen.
%
In this scheme, the output SNR can be expressed as 
$\gamma _k^{{\rm{TAS/SC}}} = \mathop {\max }\limits_{\scriptstyle i = 1,2,...,{N_T}\hfill\atop
\scriptstyle j = 1,2,...,{N_R}\hfill} \gamma _k^{\left( {i,j} \right)}$.
%
Accordingly, the CDF of $\gamma _k^{{\rm{TAS/SC}}}$ can be obtained easily as
\begin{align}
\label{Eq:CDF_TAS/SC}
{F_{\gamma _k^{{\rm{TAS/SC}}}}}\left( \gamma  \right) 
= {\left\{ {1 - \exp \left( { - \frac{\gamma }{{{{\bar \gamma }_k}}}} \right)} \right\}^{{N_T}{N_R}}}.
\end{align}

\section{Performance Analysis}
\label{Sect:PerfAnalysis}
To evaluate the system performance of SPC, we derive the closed-form expression for the e2e BLER.
The source encodes $\cal T$ information bits into a blocklength of $\beta$ CUs, where the signals are transmitted from $R_0$ to $R_{K+1}$ with the help of $R_1, R_2,...R_K$ via quasi-static fading channels \cite{yang2014quasi}.
For quasi-static fading channels, we assume that the channel fading coefficients are random, but remain constant during each transmission block and change independently at other blocks.
The coding rate of the considered system for each time slot is given by ${\cal R} = {\cal T}/\beta$.
In contrast, in the context of the FBL scheme with the SPC, the blocklength employed should be minimized, but it still needs to be longer than 100 CUs \cite{makki2014finite}. In this scenario, the maximum coding rate
can be approximately expressed as \cite{polyanskiy2010channel}  
\begin{align}
\label{Eq:MaxCodingRate}
{\cal R} \approx {\cal C}\left( {{\gamma^{\cal X} _k}} \right) - \sqrt {{\cal V}\left( {{\gamma^{\cal X} _k}} \right)/\beta } {Q^{ - 1}}\left( {{\varepsilon^{\cal X} _k}} \right),
\end{align}
where ${\cal X} \in \left\{ {{\rm{TAS/MRC}},{\rm{TAS/SC}}} \right\}$ denotes one type of the diversity transmission schemes, $\varepsilon^{\cal X} _k$ represents the instantaneous BLER at the $k$th hop, ${\cal C}\left( {{\gamma^{\cal X} _k}} \right) \buildrel \Delta \over = {\log _2}\left( {1 + {\gamma^{\cal X} _k}} \right)$ is the channel capacity, 
${\cal V}\left( {{\gamma^{\cal X} _k}} \right) \buildrel \Delta \over = \left( {1 - \frac{1}{{{{\left( {1 + {\gamma^{\cal X} _k}} \right)}^2}}}} \right){\left( {{{\log }_2}e} \right)^2}$ denotes the channel dispersion measuring the stochastic variability of the channel relative to a deterministic channel for the same capacity \cite{polyanskiy2010channel}, and 
${Q^{ - 1}}\left(  \cdot  \right)$ is the inverse function of the Q-function with $Q(z) = \frac{1}{\sqrt{2\pi}}\int_z^\infty \exp\left( -\frac{u^2}{2}\right)du$. 

From \eqref{Eq:MaxCodingRate}, the average BLER at each hop can be written as
\begin{align}
\label{Eq:AverageBLER}
{{\bar \varepsilon^{\cal X} }_k} &\approx  \int\limits_0^\infty  {Q\left( {\frac{{{\cal C}\left( {{\gamma^{\cal X} _k}} \right) - {\cal R}}}{{\sqrt {{\cal V}\left( {{\gamma^{\cal X} _k}} \right)/\beta } }}} \right){f_{{\gamma^{\cal X} _k}}}\left( \gamma  \right)d\gamma },
\end{align}
where ${f_{{\gamma^{\cal X} _k}}}\left(  \cdot  \right)$ denotes the probability density function of ${{\gamma^{\cal X} _k}}$.
Because of the complicated form of $Q\left(  \cdot  \right)$ in \eqref{Eq:AverageBLER}, the closed-form expression for the average BLER at each hop cannot be calculated directly.
Fortunately, motivated by the work in \cite[eq.~(14)]{makki2014finite}, an approximation approach of the Q-function can be leveraged to solve the problem in \eqref{Eq:AverageBLER}. In particular, we use the approximation $Q\left( {\frac{{{\cal C}\left( {{\gamma^{\cal X} _k}} \right) - {\cal R}}}{{\sqrt {{\cal V}\left( {{\gamma^{\cal X} _k}} \right)/\beta } }}} \right) \approx \Psi \left( {{\gamma^{\cal X} _k}} \right)$,
with $\Psi \left( {{\gamma^{\cal X} _k}} \right)$ represented as
\begin{align}
\label{}
\Psi \left( {{\gamma^{\cal X} _k}} \right) \approx \left\{ {\begin{array}{*{20}{l}}
{1,}&{{\gamma^{\cal X} _k} \le {\varphi _L,}}\\
{0.5 - \xi \sqrt \beta  \left( {{\gamma^{\cal X} _k} - \tau } \right),}&{{\varphi _L} < {\gamma^{\cal X} _k} < {\varphi _H,}}\\
{0,}&{{\gamma^{\cal X} _k} \ge {\varphi _H,}}
\end{array}} \right.
\end{align}
where $\xi  = {\left[ {2\pi \left( {{2^{2{\cal R}}} - 1} \right)} \right]^{ - 1/2}}$,
$\tau  = {2^{\cal R}} - 1$,
${\varphi _H} = \tau  + {1 \mathord{\left/
 {\vphantom {1 {\left( {2\xi \sqrt \beta  } \right)}}} \right.
 \kern-\nulldelimiterspace} {\left( {2\xi \sqrt \beta  } \right)}}$,
and ${\varphi _L} = \tau  - {1 \mathord{\left/
 {\vphantom {1 {\left( {2\xi \sqrt \beta  } \right)}}} \right.
 \kern-\nulldelimiterspace} {\left( {2\xi \sqrt \beta  } \right)}}$.
With this feasible approximation, the average BLER at each hop can be written as
\begin{align}
\label{Eq:AverageBLER_CDF}
{{\bar \varepsilon^{\cal X} }_k} \approx \int\limits_0^\infty  {\Psi \left( {{\gamma^{\cal X} _k}} \right){f_{{\gamma^{\cal X} _k}}}\left( {{\gamma^{\cal X} _k}} \right)d{\gamma^{\cal X} _k}} \mathop  = \limits^{\left( a \right)} \xi \sqrt \beta  \int\limits_{{\varphi _L}}^{{\varphi _H}} {{F_{{\gamma^{\cal X} _k}}}\left( \gamma  \right)d\gamma },
\end{align}
where step $\left( a \right)$ is conducted by the partial integration method.

By inserting \eqref{Eq:CDF_TAS/MRC} into \eqref{Eq:AverageBLER_CDF} and after some mathematical manipulations,
the closed-form expression of the average BLER at each hop for the TAS/MRC scheme is given by
\begin{align}
\label{eq:BLER_k_TAS/MRC_Final}
&\bar \varepsilon _k^{{\rm{TAS/MRC}}} 
\approx 1 + \xi \sqrt \beta  \sum\limits_{m = 1}^{{N_T}} {\left( {\begin{array}{*{20}{c}}
{{N_T}}\\
m
\end{array}} \right){{\left( { - 1} \right)}^m}} \nonumber\\
& \hspace{1.25cm} \cdot \sum\limits_{{j_1} = 0}^{{N_R} - 1} {\sum\limits_{{j_2} = 0}^{{N_R} - 1} { \ldots \sum\limits_{{j_m} = 0}^{{N_R} - 1} {\left( {\prod\limits_{t = 1}^m {\frac{1}{{{j_t}!}}} } \right)} } } {m^{ - {\cal S} - 1}}{{\bar \gamma }_k} \nonumber\\
& \hspace{1.25cm}\cdot \left\{ {\Upsilon \left( {{\cal S} + 1,\frac{{m{\varphi _H}}}{{{{\bar \gamma }_k}}}} \right) - \Upsilon \left( {{\cal S} + 1,\frac{{m{\varphi _L}}}{{{{\bar \gamma }_k}}}} \right)} \right\}, 
\end{align}
where ${\cal S} = \sum\nolimits_{t = 1}^m {{j_t}} $ and $\Upsilon \left( {\alpha ,z} \right) = \int_0^z {{e^{ - t}}{t^{\alpha  - 1}}dt} $ denotes the lower incomplete gamma function.

By inserting \eqref{Eq:CDF_TAS/SC} into \eqref{Eq:AverageBLER_CDF}, the average BLER at each hop for the TAS/SC scheme can be written as
\begin{align}
\label{Eq:BLER_TAS/SC1}
&\bar \varepsilon _k^{{\rm{TAS/SC}}} 
\approx 1 + \xi \sqrt \beta  \sum\limits_{m = 1}^{{N_T}{N_R}} {\left( {\begin{array}{*{20}{c}}
{{N_T}{N_R}}\\
m
\end{array}} \right){{\left( { - 1} \right)}^{m + 1}}} \nonumber\\
& \hspace{1cm} \cdot \frac{{{{\bar \gamma }_k}}}{m}\left\{ {\exp \left( { - \frac{m}{{{{\bar \gamma }_k}}}{\varphi _H}} \right) - \left\{ {\exp \left( { - \frac{m}{{{{\bar \gamma }_k}}}{\varphi _L}} \right)} \right\}} \right\}.
\end{align}

For the given average BLERs at each hop in \eqref{eq:BLER_k_TAS/MRC_Final} and \eqref{Eq:BLER_TAS/SC1}, we propose the following proposition to calculate the e2e BLERs via the SDF principle.

\begin{proposition}\label{Proposition:SDF}
Based on the SDF mechanism \cite{khafagy2013outage}, only when a relay decodes the received data correctly, it forwards the data to the next hop. 
If the incorrect decoding occurs at a relay node, the BLERs for all later nodes become one, i.e., $\bar \varepsilon _n = 1\left( {\forall n > k} \right)$, where $k$ is the index of the relay that performs erroneous decoding.
Therefore, the e2e BLER for the multihop SDF network is given by
\begin{align}\label{eq:SDF_approximate}
{{\bar \varepsilon^{\cal X} }_{{\rm{e2e}}}} 
&\approx {{\bar \varepsilon^{\cal X} }_1} + \left( {1 - {{\bar \varepsilon^{\cal X} }_1}} \right){{\bar \varepsilon^{\cal X} }_2} + ... + \left( {1 - {{\bar \varepsilon^{\cal X} }_1}} \right)...\left( {1 - {{\bar \varepsilon^{\cal X} }_K}} \right){{\bar \varepsilon^{\cal X} }_{K + 1}}\nonumber\\
&= {{\bar \varepsilon^{\cal X} }_1} + \sum\limits_{k = 2}^{K + 1} {\left( {{{\bar \varepsilon^{\cal X} }_k} \cdot \prod\limits_{m = 2}^k {\left( {1 - {{\bar \varepsilon^{\cal X} }_{m - 1}}} \right)} } \right)} .
\end{align}
\end{proposition}

\color{black}
\begin{proposition}[e2e latency and throughput]\label{Proposition:e2eLatencyThoughput}
Let ${\cal F}$ be the feedback delay in each hop, measured in CUs, and ${\cal D}\left( \beta  \right)$ denote the delay for decoding a packet with the blocklength $\beta$  \cite{makki2019fast}. In practice, ${\cal D}\left( \beta  \right)$ depends on the decoding scheme, the number of iterations in the decoder, etc. \cite{makki2019fast}.
We consider the e2e latency in the retransmission scheme, where the retransmission continues until the packet is decoded correctly or the system reaches the maximum number of retransmission times, denoted by $L$. As a result, the e2e packet transmission latency (measured in CUs) is expressed as
\begin{align}
\label{}
{\cal L}_{\rm{e2e}}^{\cal X} = \frac{{\left( {1 - \bar \varepsilon _{\rm{e2e}}^{\cal X}} \right)}}{{\left[ {1 - {{\left( {\bar \varepsilon _{\rm{e2e}}^{\cal X}} \right)}^{L + 1}}} \right]}}\left[ {\sum\limits_{r = 0}^L {{{\left( {\bar \varepsilon _{\rm{e2e}}^{\cal X}} \right)}^r}\left( {rT_F^{\cal X} + {T_S}} \right)} } \right],
\end{align}
where $\cal X$ denotes one type of transmission scheme, i.e., ${\cal X} \in \left\{ {{\rm{TAS/MRC,}}\,{\rm{TAS/SC}}} \right\}$.
Here, $T_S$ and $T_F$ represent the average latency caused by e2e decoding success and failure, respectively, given by ${T_S} = \left( {K + 1} \right)\left( {\beta  + {\cal D}\left( \beta  \right)} \right)$ and
\begin{align}
T_F^{\cal X} = \left( {\beta  + {\cal D}\left( \beta  \right) + {\cal F}} \right)\left( {\bar \varepsilon _1^{\cal X} + \sum\limits_{k = 2}^{K + 1} {k\bar \varepsilon _k^{\cal X}\cdot\prod\limits_{m = 2}^k {\left( {1 - \bar \varepsilon _{m - 1}^{\cal X}} \right)} } } \right).
\end{align}

Furthermore, the e2e throughput is defined as the ratio of the number of information bits successfully received at the destination to the average duration time, which is measured in bits per CU (BPCU).
As a result, the e2e throughput for the considered system is given by
\begin{align}
\label{}
\delta _{\rm{e2e}}^{\cal X} = \frac{{{\cal T}\left[ {1 - {{\left( {\bar \varepsilon _{\rm{e2e}}^{\cal X}} \right)}^{L + 1}}} \right]}}{{{\cal L}_{\rm{e2e}}^{\cal X}\left[ {1 - {{\left( {\bar \varepsilon _{\rm{e2e}}^{\cal X}} \right)}^{L + 1}}} \right] + \left( {L + 1} \right){T^{\cal X}_F}}{{\left( {\bar \varepsilon _{\rm{e2e}}^{\cal X}} \right)}^{L + 1}}}.
\end{align}
\end{proposition}
\color{black}

\section{Asymptotic Analysis}
\label{Sect:AsymptoticAnalysis}

We define $\bar \gamma  = {P_S}/{N_0}$ as an average SNR and assume that we allocate power for the source and relay nodes equally, i.e., ${P_{k-1}} = {P_S}/\left( {K + 1} \right)$, where $P_S$ is the total transmit power of the system.
It is noted that ${{\bar \gamma }_k} = \frac{{{P_{k - 1}}}}{{{N_0}}}\mathbb{E}\left\{ {{{\left| {h_k^{\left( {i,j} \right)}} \right|}^2}} \right\} = {c_k}\bar \gamma $, 
where ${c_k} = \mathbb{E}\left\{ {{{\left| {h_k^{\left( {i,j} \right)}} \right|}^2}} \right\}/\left( {K + 1} \right)$.
Therefore, in the high-SNR regime with $\bar \gamma  \to \infty $, we have ${{\bar \gamma }_k} \to \infty$.

In the high-SNR regime, we use
\cite{gradshteyn2014table} and
\cite{duong2012cognitive} for \eqref{Eq:AverageBLER_CDF}. The asymptotic BLER at each hop for TAS/MRC can be given by
\begin{align}\label{eq:26}
\tilde \varepsilon _k^{\infty {\rm{TAS/MRC}}} 
= \frac{{\xi \sqrt \beta  \left[ {\varphi _H^{{N_T}{N_R} + 1} - \varphi _L^{{N_T}{N_R} + 1}} \right]}}{{{{\left( {{N_R}!} \right)}^{{N_T}}}\bar \gamma _k^{{N_T}{N_R}}\left( {{N_T}{N_R} + 1} \right)}}.
\end{align}
For TAS/SC, when ${\bar\gamma _{k}} \to \infty $, we utilize $1 - \exp \left( { - \frac{\gamma }{{{{\bar \gamma }_k}}}} \right)\ \sim \frac{\gamma }{{{{\bar \gamma }_k}}}$ for \eqref{Eq:CDF_TAS/SC}. The asymptotic BLER at each hop for the TAS/SC scheme is expressed as
\begin{align}\label{eq:28}
\tilde \varepsilon _k^{\infty {\rm{TAS/SC}}} 
= \frac{{\xi \sqrt \beta  \left[ {\varphi _H^{{N_T}{N_R} + 1} - \varphi _L^{{N_T}{N_R} + 1}} \right]}}{{\bar \gamma _k^{{N_T}{N_R}}\left( {{N_T}{N_R} + 1} \right)}}.
\end{align}

In the same manner as \textit{Proposition~\ref{Proposition:SDF}}, we invoke \eqref{eq:SDF_approximate} for the average BLER at each hop such that given in \eqref{eq:26} and \eqref{eq:28}. 
The value of $\tilde \varepsilon _k^{\infty \cal X} $ is very small for ${{\bar \gamma }_k} \to \infty$, i.e., $\tilde \varepsilon _k^{\infty \cal X} \ll 1$. Therefore, the asymptotic e2e BLER can be expressed approximately as $\tilde \varepsilon _{{\rm{e2e}}}^{\infty \cal X} \approx \sum\limits_{k = 1}^{K + 1} {\tilde \varepsilon _k^{\infty \cal X} }$.

\begin{proposition}[e2e diversity order, BLER loss, and SNR gap]\label{Corollary:DiversityOrder}
For the diversity order in the form of
%
\begin{align}\label{eq:diversity_order}
{D}_{\cal X} =  - \mathop {\lim }\limits_{\bar \gamma  \to \infty } \frac{{\log \left( {\tilde \varepsilon _{{\rm{e2e}}}^{\infty {\cal X}}} \right)}}{{\log \left( {\bar \gamma } \right)}},
\end{align}
the maximum diversity orders for both the TAS/MRC and TAS/SC schemes are achieved, i.e., ${D}_{\cal X} = N_T N_R$.
For the BLER loss, we consider the ratio of the asymptotic e2e BLER for TAS/MRC to the asymptotic e2e BLER for TAS/SC as
\begin{align}\label{eq:BLER_loss}
\frac{{\tilde \varepsilon _{{\rm{e2e}}}^{\infty {\rm{TAS/MRC}}}}}{{\tilde \varepsilon _{{\rm{e2e}}}^{\infty {\rm{TAS/SC}}}}} 
\approx \frac{{\sum\limits_{k = 1}^{K + 1} {\tilde \varepsilon _k^{\infty {\rm{TAS/MRC}}}} }}{{\sum\limits_{k = 1}^{K + 1} {\tilde \varepsilon _k^{\infty {\rm{TAS/SC}}}} }} 
= \frac{1}{{{{\left( {{N_R}!} \right)}^{{N_T}}}}}.
\end{align}
To gain further insight, we can rewrite the asymptotic e2e BLER of $\cal X$ as $\tilde \varepsilon _{\rm{e2e}}^{\infty {\cal X}} \approx {\left( {{{\cal G}_{\cal X}}\bar \gamma } \right)^{ - {D_{\cal X}}}}$, 
where 
${{\cal G}_{\cal X}} = {\left( {{{\cal Y}_{\cal X}}\sum\limits_{k = 1}^{K + 1} {\frac{1}{{c_k^{{N_T}{N_R}}}}} } \right)^{ - \frac{1}{{{D_{\cal X}}}}}}$ 
is the array gain, which represents the SNR gain in the e2e BLER curve with respect to ${{\bar \gamma }^{ - {D_{\cal X}}}}$. Here, ${{\cal Y}_{\cal X}}$ is defined as
\begin{align}
\label{}
{{\cal Y}_{\cal X}} = \left\{ {\begin{array}{*{20}{l}}
{\frac{{\xi \sqrt \beta  \left( {\varphi _H^{{N_T}{N_R} + 1} - \varphi _L^{{N_T}{N_R} + 1}} \right)}}{{{{\left( {{N_R}!} \right)}^{{N_T}}}\left( {{N_T}{N_R} + 1} \right)}},}&{{\rm{if}}\,{\cal X} = {\rm{TAS/MRC,}}}\\
{\frac{{\xi \sqrt \beta  \left( {\varphi _H^{{N_T}{N_R} + 1} - \varphi _L^{{N_T}{N_R} + 1}} \right)}}{{\left( {{N_T}{N_R} + 1} \right)}},}&{{\rm{if}}\,{\cal X} = {\rm{TAS/SC.}}}
\end{array}} \right.
\end{align}
Accordingly, the SNR gap between TAS/MRC and TAS/SC is represented by the ratio of their array gains,
given by
\begin{align}
\label{eq:36}
\frac{{{{\cal G}_{{\rm{TAS/MRC}}}}}}{{{{\cal G}_{{\rm{TAS/SC}}}}}} = {\left[ {\frac{{{{\cal Y}_{{\rm{TAS/SC}}}}}}{{{{\cal Y}_{{\rm{TAS/MRC}}}}}}} \right]^{\frac{1}{{{N_R}{N_T}}}}} = {\left( {{N_R}!} \right)^{\frac{1}{{{N_R}}}}}.
\end{align}
\end{proposition}

\begin{remark}\label{Remark:1}
Although TAS/MRC and TAS/SC offer the same full diversity order, they have different array gains, which leads to the e2e BLER loss and SNR gap, as described in \eqref{eq:BLER_loss} and \eqref{eq:36}, respectively.
\end{remark}

\section{Simulation Results and Discussion}
\label{Sect:SimuDiscuss}

In this section, we have performed Monte Carlo simulations to validate our theoretical results and evaluate the e2e performance of the proposed system.
For channel settings, we consider the simplified path-loss model \cite{goldsmith2005wireless} with the average channel gain ${\mathbb{E}}\left\{ {{{\left| {h_k^{\left( {i,j} \right)}} \right|}^2}} \right\} = d_k^{ - \eta }$, 
where $d_k$ is the distance between two adjacent nodes and $\eta = 3$ denotes the path-loss exponent.
Assuming that we employ equal allocation for both the power and relay location configurations, e.g., 
${d_k} = D/\left( {K + 1} \right)$ and ${P_{k-1}} = P_{S}/\left( {K + 1} \right)$,
where $D = 1$ is the normalized transmission distance.

\begin{figure}[!t]
\centering
\includegraphics[width=0.7\linewidth]{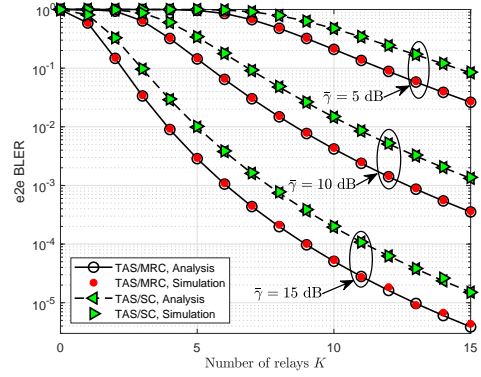}
\caption{e2e BLER versus the number of relays $K$, where $N_T = N_R = 2$, $\mathcal{T}  = 1024$ bits, and $\beta=128$ CUs.}
\label{Fig:2}
\end{figure}

Fig.~\ref{Fig:2} presents the influence of the number of relays $K$ on the system performance. From Fig.~\ref{Fig:2}, we make various observations, including (\textit{i}) TAS/MRC vs. TAS/SC performance, (\textit{ii}) the influence of the number of relays $K$, and (\textit{iii}) the effect of the average SNR.
First, it is clear that the TAS/MRC scheme achieves better performance than the TAS/SC over the entire SNR range.
However, it is worth noting that TAS/SC requires fewer RF chains than TAS/MRC, which leads to lower power consumption and hardware complexity.
From the second perspective, the gain of multihop employment in terms of e2e performance is clearly shown.
The e2e BLER dramatically drops when the number of relays $K$ is increased.
This property can be explained by the fact that when $K$ increases, $d_k$ decreases, which reduces the BLER at each hop, resulting in a reduction in e2e BLER.
In Fig.~\ref{Fig:2}, we also observe the e2e BLER performance versus $\bar \gamma$.
It is seen that as $\bar \gamma$ increases, the performance is improved.

\begin{figure}[!t]
\centering
\includegraphics[width=0.7\linewidth]{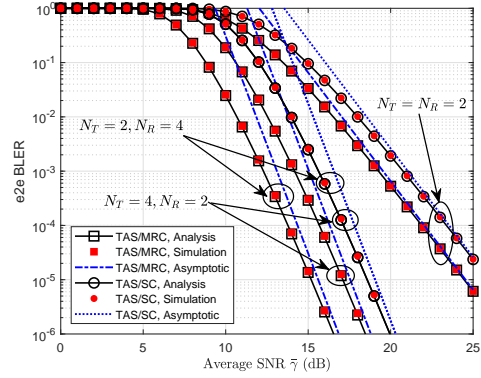}
\caption{The comparison for various MIMO configurations, where $K=3$, $\mathcal{T} = 1024$ bits, and $\beta=128$ CUs.}
\label{Fig:3}
\end{figure}

In Fig.~\ref{Fig:3}, we investigate the system performance versus the average SNR for various MIMO configurations.
Fig.~\ref{Fig:3} shows that, as expected, better performance is achieved by employing more antennas, owing to higher diversity gains.
It is also observed that for the two scenarios of $N_T = 2, N_R = 4$ and $N_T = 4, N_R = 2$, the TAS/SC has identical performance, whereas the performance of the TAS/MRC scheme varies.
This is because the TAS/SC scheme chooses only one link out of a set of $N_T \times N_R$ transmit/receive antenna combinations, and the size of this set is the same for both scenarios, which leads to the same e2e BLER performance. In contrast, for the TAS/MRC scheme, the MRC is employed at the receiver, and hence the employment of more receive antennas is more beneficial than that of more transmit antennas, which results in better performance for $\left( {{N_T} = 2,{N_R} = 4} \right)$ than for $\left( {{N_T} = 4,{N_R} = 2} \right)$.
%

%

In Figs.~\ref{Fig:2} and \ref{Fig:3}, we recognize that the simulation results are in good agreement with all the analysis results, which verifies our analytical correctness.
Additionally, the analysis and simulation results converge to the asymptotic results at high SNRs in Fig.~\ref{Fig:3}.
It is worth noted that as the SNR increases, the e2e BLER loss between the TAS/MRC and TAS/SC schemes converges to ${\left( {{N_R}!} \right)^{{N_T}}}$ and the SNR gap between them converges to $10\log \left( {{{\left( {{N_R}!} \right)}^{\frac{1}{{{N_R}}}}}} \right)$ in dB, as presented in \textit{Proposition~\ref{Corollary:DiversityOrder}}.



\begin{figure}[!t]
\centering
\includegraphics[width=0.7\linewidth]{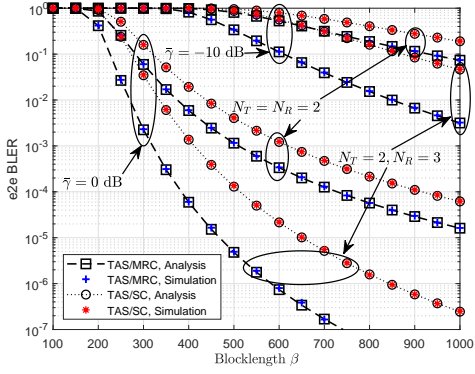}
\caption{e2e BLER versus $\beta$, where $K=3$ and $\mathcal{T} = 1024$ bits.}
\label{Fig:5}
\end{figure}

To observe the impact of the blocklength $\beta$ on the system performance, we plot the e2e BLER as a function of blocklength $\beta$ in Fig.~\ref{Fig:5}.
Fig.~\ref{Fig:5} shows that the increase in $\beta$ leads to better system performance. 
We note that the main purpose of SPC is to maintain the blocklength as short as possible but it still employs more than 100 CUs to achieve both low latency and reliable communications.
Considering this tradeoff, the value of $\beta$ is chosen conservatively to satisfy both the minimum-latency constraint and BLER performance.
For example, let us assume $N_T = 2, N_R = 3$, an average SNR of $0$ dB, and quality of service with a BLER of $10^{-5}$.
In this scenario, $\beta$ can be chosen as $\beta \approx 460$ and $\beta \approx 650$ CUs for the TAS/MRC and TAS/SC schemes, respectively, which implies that our analysis results for e2e BLER, presented in Section~\ref{Sect:PerfAnalysis}, can be employed to optimize the SPC system.


\begin{figure}[!t]
\centering
\includegraphics[width=0.7\linewidth]{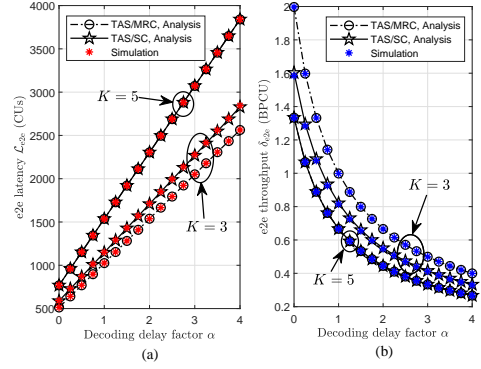}
\caption{The e2e latency and throughput versus the decoding delay factor $\alpha$, where $N_T = N_R = 3$, $\bar\gamma = 10$~dB, $\mathcal{T} = 1024$ bits, $L = 20$, ${\cal F} = 40$ CUs, and $\beta = 128$ CUs.}
\label{Fig:6}
\end{figure}

To evaluate the e2e latency and throughput given in \textit{Proposition~\ref{Proposition:e2eLatencyThoughput}}, we consider a linear decoding delay profile ${\cal D}\left( \beta  \right) = \alpha \beta $, where $\alpha$ is the constant decoding delay factor \cite{makki2019fast}, which depends on different decoding schemes.
In Fig.~\ref{Fig:6}, for various decoding delay factors, the e2e performance is compared in terms of both the e2e latency and throughput.
We also observe that TAS/MRC achieves lower latency and higher throughput compared to TAS/SC when $K=3$.
In contrast, for $K=5$, the decoding error probability is very close to zero in both the TAS/MRC and TAS/SC schemes; hence, they have almost the same latency and throughput.


\begin{figure}[!t]
\centering
\includegraphics[width=0.7\linewidth]{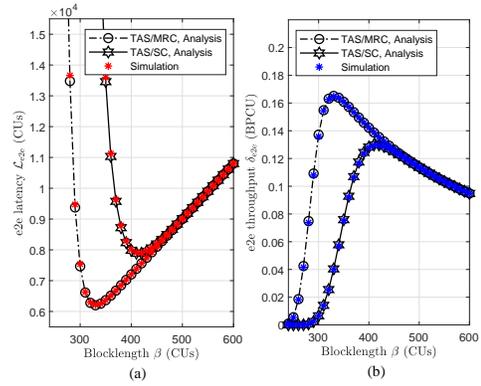}
\caption{The e2e latency and throughput versus the blocklength $\beta$, where $N_T = N_R = 3$, $K=5$, $\bar\gamma = -10$~dB, $\alpha = 2$, $L = 20$, ${\cal F} = 40$ CUs, and $\mathcal{T} = 1024$ bits.}
\label{Fig:7}
\end{figure}

Fig.~\ref{Fig:7} depicts the e2e latency and throughput of the considered systems for various blocklengths $\beta$.
In Fig.~\ref{Fig:7}, very high e2e latency and very low e2e throughput were attained for short blocklengths.
It is due to the fact that, with the short blocklength,
the e2e BLER is high, which leads to the more frequent restransmissions. 
In contrast, when the blocklength is sufficiently long, the probability of retransmitting the packet converges to zero.
In this scenario, increasing $\beta$ leads to an increase in the e2e latency and a decrease in the e2e throughput.
Therefore, choosing the optimal value for $\beta$ is worth considering.
In Fig.~\ref{Fig:7}(a) and Fig.~\ref{Fig:7}(b), it can be seen that for a given number of information bits $\cal T$, $\beta = 330$ CUs and $\beta = 420$ CUs are the optimal blocklength values for the TAS/MRC and TAS/SC schemes, respectively, which minimize the e2e latency and maximize the e2e throughput.



\begin{figure}[!t]
\centering
\includegraphics[width=0.7\linewidth]{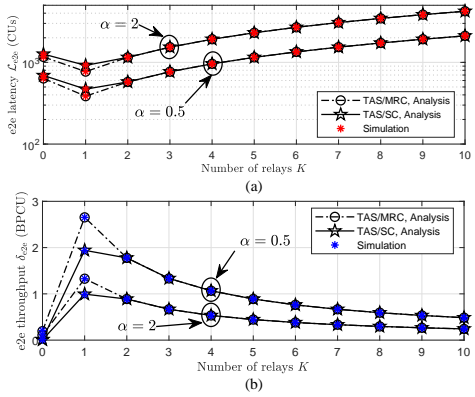}
\caption{The influence of the number of relays $K$ on the e2e latency and throughput, where $N_T = N_R = 3$, $\bar\gamma = 15$~dB, $\mathcal{T} = 1024$ bits, $L = 20$, ${\cal F} = 40$ CUs, and $\beta = 128$ CUs.}
\label{Fig:8}
\end{figure}

In multihop relay networks for uRLLCs, the number of relay nodes plays an important role.
Fig.~\ref{Fig:8} shows the influence of $K$ on the e2e latency and throughput performance.
In Fig.~\ref{Fig:8}, it is observed that the e2e latency and throughput are minimized and maximized for $K=1$, respectively.
Intuitively, a larger number of relays leads to shorter hops, i.e., shorter distance between two relay nodes to improve the reliability, but it can suffer from a longer transmission time via relays and even e2e throughput degradation.
Nonetheless, under the tradeoff consideration among e2e BLER, latency, and throughput, we can determine the optimal value of the number of relays.
For example, for the environment shown in Fig.~\ref{Fig:8}, $K=1$ is the optimal value to achieve the highest e2e throughput and lowest e2e latency for both the TAS/MRC and TAS/SC schemes. 
For a CU duration of 3~${\mu }$s \cite{lopez2017ultrareliable} and $\alpha  = 2$, the e2e latencies for TAS/MRC and TAS/SC are 770 and 1030 CUs corresponding to 2.31 and 3.09~ms, respectively, which satisfy the low-latency constraint of uRLLCs (${ \le 10}$~ms \cite{lopez2017ultrareliable}).

\color{black}

\section{Conclusions}
\label{Sect:Conclusion}
In this work, we have investigated the MIMO SDF multihop relay network for the SPC scenario.
Under the consideration of the FBL regime and quasi-static Rayleigh fading channels, the closed-form expressions for the e2e BLERs of the TAS/MRC and TAS/SC schemes are obtained in approximated forms, from which their asymptotic forms for high SNRs are derived.
Based on the asymptotic analysis, the e2e diversity order, e2e BLER loss, \textcolor{black}{and SNR gap} are analyzed. Furthermore, the e2e latency and throughputs are also analyzed for the considered schemes. The simulation results demonstrate the gains from the employment of MIMO and multihop relay network for uRLLC requirements using SPC.


\bibliographystyle{IEEEtran}
\bibliography{references}
\end{document}